# Nucleus-nucleus reaction cross-sections for deformed nuclei

**M.Y.M. Hassan**[1], **M.Y.H. Farag**[1], **A.Y. Abul-Magd**[2] and
**T.E.I. Nassar**[3]

[1] Physics Department, Faculty of Science, Cairo University, Cairo, Egypt.
[2] Faculty of Engineering, Sinai University, El-Arish, Egypt.
[3] Mathematics Department, Faculty of Science, Zagazig University, Zagazig, Egypt.

**Abstract**

Reaction cross-sections are calculated using the Coulomb modified Glauber model for deformed target nuclei. The deformed nuclear matter density of the target is expanded into multipoles of order $k = 0, 2, 4$. The reaction cross-section for $^{12}C + ^{27}Al$, $^{20}Ne + ^{27}Al$, $^{12}C + ^{64}Zn$, $^{12}C + ^{90}Zr$, $^{40}Ar + ^{238}U$ and $^{20}Ne + ^{235}U$ are studied at energy range (10-1000 MeV/nucleon). The most significant effects in the intermediate energy range are the Coulomb field and in-medium effect that modified the trajectory of the incident beams. Introducing the deformation effect beside the Coulomb field and in-medium effect improves the agreement with the experimental data and two empirical parameterizations in the case of not finding experimental data. Moreover it is indicated that the enhancement of the reaction cross-sections is attributed with fixed orientation in deformed nuclei.

# [1] Introduction:

In recent years, an increasing number of both theoretical [1-5] and experimental [6-10] calculations have been performed to study the reaction cross-section for nucleus-nucleus collision. The reaction cross-section is one of the most important physical quantities required to characterize the nuclear reaction. It is very useful for extracting nuclear size and it finds other applications in various fields of research as radiobiology and space radiation [11, 12]. The Glauber model has been commonly used to describe heavy ion reactions at high energies. It is a semi-classical model picturing the projectile beams traversing the target nucleus in a straight path along the incident particles direction, which is mainly based on the independent nucleon-nucleon collision in the overlap zone of the colliding nuclei [1]. The Glauber approach provides reasonable agreement with experiment at high energies. Karol [4], used Glauber model for calculating the reaction cross-section at 2.1 GeV/nucl for spherical nuclei, where Coulomb effect, Fermi motion and Pauli blocking etc., were ignored. The model gives good result at high energy. While, the Glauber model failed to describe the lower energy collisions, this disagreement is due to the significant role played by the Coulomb potential, whose effects are obvious in low and intermediate energies. The optical limit approximation to Glauber's model was modified to take into account the Coulomb distortion of the straight line trajectory occurring in the case of heavy ion scattering at low energies [13-17]. This modification is called the Coulomb modified Glauber and denoted by " Glauber model I". The modified Glauber model I was applied for many spherical nucleus-nucleus interactions [13-20]. All of the $^{12}C+^{27}Al$, $^{20}Ne+^{27}Al$ $^{12}C+^{64}Zn$ and $^{12}C+^{90}Zr$ reactions were calculated with different methods at specific energies [2,5,9,21-23]. The reaction cross-section for these reactions was calculated considering spherical target nuclei, while, in our research we considered the deformation of the target nuclei to calculate the reaction cross-sections. [2, 3] For nuclei which have no experimental data, the results are compared with empirical parameterization of the reaction cross-section. Both of the two reactions $^{40}Ar+^{238}U$ and $^{20}Ne+^{235}U$ were calculated by the empirical parameterization [2] at different energies, since they have little experimental data at low energies.

In the present work, the modified Glauber model I is applied to calculate the reaction cross-sections for deformed target nuclei with deformation parameter $\beta_2$. In these investigations, the zero range nucleon-nucleon (NN) amplitude, with energy range from 10-1000 MeV/nucleon is assumed. The nucleon-nucleon amplitude is modified by introducing the in-medium effect [24-26]. We found that the reaction cross-section changes significantly during orientation of the deformed nucleus. The paper is organized as follows. Section 2 contains the theory, while the calculations are presented in section 3. Discussions and conclusions are given in section 4.

## [2] Theory:

The Glauber model has been commonly used to calculate the high energy reaction cross sections for nucleus-nucleus collisions. In the Glauber model, the nuclear phase shift of nucleus–nucleus collision is the sum of nucleon-nucleon phase shift of nucleons in the two nuclei [1]. The nuclear phase shift is considered for the two colliding nuclei, such that one of them is deformed [27-30]. In the present work the quadrupole deformation and modified Glauber model are studied with fixed orientations to calculate the reaction cross-sections. Also, the in-medium effect is studied.

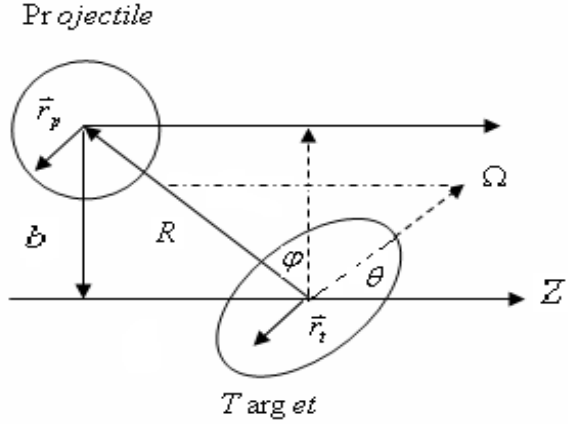

**Figure 1**: Schematic representation of the collision of the spherical projectile and quadrupole deformed target nucleus at an impact parameter $\vec{b}$.

At high energies the optical limit reaction cross-section $\sigma_R$ can be written as a function of deformation parameter $\beta_2$ and a fixed orientation $\hat{\Omega}$ of the incident spherical projectile and the deformed target nucleus as [28].

$$\sigma_R(\beta_2,\hat{\Omega}) = 2\pi \int_0^\infty db\, b[1-T(b,\hat{\Omega})]. \tag{1}$$

Here $T(b,\hat{\Omega})$ is the transparency function "the probability of the projectile to traverse the target nucleus without interactions at center of mass impact parameter $b$". It is given as:

$$T(b,\hat{\Omega}) = \exp(-\sigma_{NN}\chi(b,\hat{\Omega})), \tag{2}$$

where $\sigma_{NN}$ is the average nucleon-nucleon cross section, while the z-integral on the overlap of the colliding two nuclei "$\chi(b,\hat{\Omega})$", is given by [28]

$$\chi(b,\hat{\Omega}) = \int_{-\infty}^{\infty} dz \int d\vec{r}_t\, \rho^{(p)}(|\vec{R}+\vec{r}_t|)\rho^{(t)}(\vec{r}_t,\hat{\Omega}), \tag{3}$$

where the separation distance between projectile and target centers is $\vec{R} = \vec{b} + \hat{k}z$. The nuclear density of target nucleus is considered as a Woods-Saxon form factor with a quadrupole deformed radius parameter of mass number $A_t$. This is given [27] as:

$$\rho^{(t)}(\vec{r},\hat{\Omega}) = \frac{\rho_0^{(t)}}{(1+\exp[(r-R(\hat{\Omega}))/a])} \quad, \tag{4}$$

assuming a fixed value for the diffuseness parameter $a = 0.564$ fm [28]. In the case of pure quadrupole deformation, the nuclear surface radius is defined as:

$$R(\hat{\Omega}) = R_0(1+\beta_2 Y_{20}(\theta,\phi)), \tag{5}$$

in which $R_0$ is the radius of comparable sphere. The method is applied to axially coplanar nuclei. The deformed Woods-Saxon form can be expanded in powers of the quadrupole deformation parameter $\beta_2$, it can be written as [27]:

$$\rho^{(t)}(\vec{r},\hat{\Omega}) = \sum_k \rho_k^{(t)}(r) P_k(\hat{r}.\hat{\Omega}) . \tag{6}$$

The magnitudes of densities $\rho_0(r)$, $\rho_2(r)$ and $\rho_4(r)$ are obtained in terms of $\beta_2$ up to $\beta_2^3$ as given in [27]:

$$\rho_0^{(t)}(r) = \rho_0^{(t)}[E_0 + \frac{\overline{\beta}_2^2}{10}E_2 + \frac{\overline{\beta}_2^3}{105}E_3], \tag{7a}$$

$$\rho_2^{(t)}(r) = \rho_0^{(t)}[\overline{\beta}_2 E_1 + \frac{\overline{\beta}_2^2}{7}E_2 + \frac{\overline{\beta}_2^3}{14}E_3], \tag{7b}$$

$$\rho_4^{(t)}(r) = \rho_0^{(t)}[\frac{9\overline{\beta}_2^2}{35}E_2 + \frac{18\overline{\beta}_2^3}{385}E_3], \tag{7c}$$

such that, $\overline{\beta}_2 = R_0\sqrt{\frac{2k+1}{4\pi}}\beta_2$,

$$E_0 = \frac{1}{1+\exp(\frac{r-R_0}{a})} \quad \text{and} \quad E_n = (-)^n \frac{d^n}{dr^n}E_0 \, , \text{ for } n=1,2,3,... \, . \tag{8}$$

The Legendre polynomial function $P_k(\hat{r}.\hat{\Omega})$ is considered for multipoles of order $k = 0,2,4$ as:

$$P_k(\hat{r}.\hat{\Omega}) = \frac{4\pi}{2k+1}\sum_{m=-k}^{k} Y_{km}^*(\hat{r})Y_{km}(\hat{\Omega}) . \tag{9}$$

The parameters of the Woods-Saxon form are determined by the volume integral and the root mean square radius [14] of the nucleus as:

$$A_t = 4\pi \int dr\, r^2 \rho_0^{(t)}(r) \quad \text{and} \quad \langle r^2 \rangle = \frac{4\pi}{A_t} \int dr\, r^4 \rho_0^{(t)}(r), \tag{10}$$

therefore

$$\rho_0^{(t)} = \frac{3A_t}{4\pi R_0^3 (1+(\frac{\pi a}{R_0})^2)} \quad \text{and} \quad R_0 = \sqrt{\frac{(5\langle r^2 \rangle - 7(\pi a)^2)}{3}}. \tag{11}$$

The density of the spherical projectile nucleus of mass $A_p$ is assumed in a Gaussian form as [28]:

$$\rho^{(p)}(|\vec{R}+\vec{r}|) = \rho_0^{(p)} \exp(-\gamma |\vec{R}+\vec{r}|^2), \tag{12}$$

With a strength $\rho_0^{(p)}$ and inverse range $\gamma$ determined by $A_p$ and diffuseness parameter $a_p = \sqrt{\frac{2}{3}} \langle r^2 \rangle^{\frac{1}{2}}$, where $\langle r^2 \rangle^{\frac{1}{2}}$ is the root mean square radius [31]. They may be expressed as:

$$\rho_0^{(p)}(0) = \frac{A_p}{(a_p \sqrt{\pi})^3} \quad \text{and} \quad \gamma = \frac{1}{a_p^2}. \tag{13}$$

For $R = \sqrt{b^2 + z^2}$ Eq. (12) can be written as:

$$\rho^{(p)}(|\vec{R}+\vec{r}|) = \rho_0^{(p)} \exp(-\gamma R^2) \exp(-\gamma r^2) \exp(-2\gamma \vec{R}.\vec{r}). \tag{14}$$

Where $\exp(-2\gamma \vec{R}.\vec{r})$ can be expressed as:

$$\exp(i(2i\gamma \vec{R}.\vec{r})) = \sum_k (2k+1) i^k j_k (2i\gamma R r) P_k(\hat{R}.\hat{r}). \tag{15}$$

Here $j_k$ is the spherical Bessel Therefore, the spherical Gaussian density of the projectile Eq. (12) is expressed as:

$$\rho^{(p)}(|\vec{R}+\vec{r}|) = \rho_0^{(p)} \exp(-\gamma R^2)\exp(-\gamma r^2) \times$$
$$4\pi \sum_{k,m} i^k j_k(2i\gamma Rr) Y_{km}^*(\hat{r}) Y_{km}(\hat{R}) . \qquad (16)$$

Using Eqs. (6), (9), (16) and the properties of spherical harmonic functions, the z-integral on the overlap of the colliding two nuclei for even multiples of order $k = 0,2,4$ is

$$\chi(b,\hat{\Omega}) = 4\pi \exp(-\gamma b^2) \sum_{k=0,2,4} \int_{-\infty}^{\infty} dz \rho_0^{(p)} \exp(-\gamma z^2) \{\int_0^{\infty} r^2 dr \rho_k^{(t)}(r)$$
$$\times \exp(-\gamma r^2) i^k j_k(2i\gamma Rr)\} P_k(\hat{R}.\hat{\Omega}) . \qquad (17)$$

If $\hat{\Omega}$ is expressed in terms of spherical polar angles $(\theta, \phi)$, with respect to the z-axis in the projectile beam direction and the center of mass impact parameter $b$, then:

$$\hat{R}\cdot\hat{\Omega} = \frac{(b\sin\theta\cos\phi + z\cos\theta)}{R} . \qquad (18)$$

The nucleon-nucleon amplitude can be modified by introducing the in-medium effect. This method includes several renormalization effects, which are difficult to calculate in a microscopic theory. In this case, the nucleon-nucleon amplitude is given in [24] as:

$$f_{NN}^{Medium} = f_M f_{NN}^{Free} , \qquad (19)$$

where $f_{NN}^{Medium}$ is the nucleon-nucleon amplitude in the medium, $f_{NN}^{Free}$ is the free nucleon-nucleon amplitude and $f_M$ represents the system energy dependent function. Then, the nucleon-nucleon cross section in the medium "$\sigma_{NN}^{Medium}$" can be represented as in [24-26], using Eq. (19) as:

$$\sigma_{NN}^{Medium} = f_M \sigma_{NN}^{Free} , \qquad (20)$$

$\sigma_{NN}^{free}$ is the nucleon-nucleon cross section in free space and $f_M$ is the renormalized amplitude which is extracted directly from the experiment and is a reliable measure of the medium modifications. The medium multipliers $f_M$ defined in Eq. (20) for ion kinetic energy in laboratory system is written as follows:

$$f_M = 0.1\exp(-\frac{E}{H}) + [1 - (\frac{\rho_{av}}{T})^{\frac{1}{3}} \exp(-\frac{E}{D})] , \qquad (21)$$

for $A_T < 56$

$$D = 46.72 + 2.21 A_T - 0.0225 A_T^2 \quad (MeV), \quad (22)$$

but for $A_T \geq 56$

$$D = 100 \quad (MeV), \quad (23)$$

where $H = 12$ $MeV$, $T = 0.14$ $fm^{-3}$ and the parameter $E$ is the laboratory energy in $MeV$, and $\rho_{av}$ refers to the average density of colliding nuclei as:

$$\rho_{av} = \frac{1}{2}(\rho_{A_p} + \rho_{A_t}), \quad (24)$$

where the density of a nucleus $A_i$ ($i = p, t$) is calculated in the hard-sphere model and is given by [25].

$$\rho_i = \frac{3A_i}{4\pi r_i^3}, \quad (25)$$

for

$$r_i = 1.29 \langle r_i^2 \rangle^{\frac{1}{2}}, \quad (26)$$

where $\langle r_i^2 \rangle^{\frac{1}{2}}$ is the root mean square radius [31]. The average nucleon-nucleon cross section is defined as [14, 31]:

$$\sigma_{NN} = \frac{(Z_P Z_T + N_P N_T)\sigma_{PP} + (Z_P N_T + N_T Z_P)\sigma_{np}}{A_T A_P}, \quad (27)$$

where $Z_P$, $Z_T$, $N_P$, $N_T$, $A_P$ and $A_T$ are the charge, neutron, and mass numbers, of the projectile and target ,respectively. Here, $\sigma_{nn}$ (or $\sigma_{pp}$) is the neutron-neutron (or proton-proton) cross section and $\sigma_{np}$ is neutron-proton cross section. The nucleon-nucleon cross sections $\sigma_{NN}$ and the ratio $\alpha_{NN}$ at different energies are given in table [1].

**Table [1]:** Parameters of the nucleon-nucleon cross section for different energies per nucleon as Refs. [14, 31].

| $E_A$ $(MeVA^{-1})$ | 10.25 | 25 | 30 | 40 | 85 | 120 | 200 | 342.5 | 550 | 800 | 1000 |
|---|---|---|---|---|---|---|---|---|---|---|---|
| $\sigma_{NN}$ ($fm^2$) | 78.7 | 24.1 | 19.6 | 13.5 | 6.1 | 4.5 | 3.2 | 2.84 | 3.62 | 4.26 | 4.32 |

The deformation parameter of any deformed nucleus can be determined as given in [32]. In this work the quadrupole deformation parameter $\beta_2$ for $^{235}U$, $^{238}U$, $^{64}Zn$, $^{27}Al$ and $^{90}Zr$ are given in table [2].

At lower energy, the optical limit form of the reaction cross-section has been modified taking into account the Coulomb field effect. The straight line trajectory is modified due to the Coulomb field between two colliding nuclei. The Coulomb modified Glauber model (MGM I) [13-15], consists in replacing the eikonal trajectory at impact parameter $b$ with the eikonal trajectory at the corresponding distance $b_c$ of closest approach in the presence of the Coulomb field. Several attempts have been mad to include the Coulomb effect into the Glauber formalism [13-20]. The reaction cross-section for deformed and Coulomb modified Glauber model can be written as:

$$\sigma_R^{ModI}(\beta_2,\hat{\Omega}) = 2\pi \int_0^\infty db\, b[1 - T(b_c,\hat{\Omega})] \ , \tag{28}$$

the value of $b_c$ is related to the impact parameter $b$ by the relation [13]:

$$b_c = \frac{1}{k}(\eta + \sqrt{\eta^2 + k^2 b^2}) \ , \tag{29}$$

where

$$\eta = \frac{Z_P Z_T e^2}{\hbar v} \ . \tag{30}$$

$k$ and $\eta$ are the projectile wave number and the Sommerfeld parameter, respectively. The relative velocity of the two nuclei is $v$. The modified deformed reaction cross-section can be written in the form [18]:

$$\bar{\sigma}_R^{ModI}(\beta_2,\hat{\Omega}) = 2\pi(1-\frac{V_C}{E_{c.m}})\int_0^\infty db\, b[1-T(b,\hat{\Omega})] \ , \qquad (31)$$

where

$$V_C = \frac{Z_T Z_P e^2}{R_C} \ . \qquad (32)$$

The strong absorption radius [20] is replaced by Coulomb radius $R_C = 1.5(A_P^{\frac{1}{3}} + A_T^{\frac{1}{3}})$ [18], to compute the Coulomb potential $V_C$.

**Table [2]:** Prolate quadrupole deformation parameter $\beta_2$ for $^{235}U$, $^{238}U$, $^{64}Zn$, $^{27}Al$ and $^{90}Zr$ Ref. [33].

| Nucleus | $^{235}U$ | $^{238}U$ | $^{27}Al$ | $^{64}Zn$ | $^{90}Zr$ |
|---|---|---|---|---|---|
| $\beta_2$ | 0.215 | 0.215 | 0.448 | 0.219 | 0.035 |

## [3] Calculations and results:

In the present work, the reaction cross-section for different deformed target nuclei is calculated with the zero range average nucleon-nucleon cross section. The reactions $^{12}C+^{27}Al$, $^{20}Ne+^{27}Al$, $^{12}C+^{64}Zn$, $^{12}C+^{90}Zr$ $^{20}Ne+^{235}U$, and $^{40}Ar+^{238}U$ are studied in the energy range 10–1000 MeV/nucleon. The measurements of "$\sigma_R$" represent a unique way of collecting information about the nuclear size of unstable nuclei. Most of the reactions have experimental data at different energies [2,5,9,22,23]. But, the reactions $^{20}Ne+^{235}U$ and $^{40}Ar+^{238}U$ have no experimental data to compare with at large energy, therefore we used two formulae [2,3] that are extracted from the reaction cross-sections for different nuclei but at incident energies from 30 to 1000 MeV/nucl. In this case, we compared our results with these empirical formulas [2, 3].

The reaction cross sections for the $^{12}C+^{27}Al$ interaction at energy range from 10 to1000 MeV/nucl, are calculated by using the modified Glauber model I with in-medium effect and prolate quadrupole deformation parameter $\beta_2$=0.448 table [2] are presented in Fig.2. The experimental data are taken from [2, 23], and the results of two empirical formulas are listed in table [3]. We notice that the reaction cross-sections that are calculated with Eq. (3) in Ref. [2] differ slightly from the ones that are calculated in Ref. [3] with Eq. (13).

**Table [3]:** The reaction cross-sections for $^{12}C + ^{27}Al$ at different energies, using two formulae Refs. [2, 3] compared with experimental data in Refs. [2,23].

| $E_A$ (MeV/nucl) | 30 | 40 | 83 | 200 | 250 | 300 | 900 |
|---|---|---|---|---|---|---|---|
| $\sigma_R$ (mb) Ref.[2] | 1711 | 1617 | 1376 | 1223 | 1217 | 1225 | 1317 |
| $\sigma_R$ (mb) Ref.[3] | 1761 | 1523 | 1276 | 1198 | 1200 | 1202 | 1272 |
| $\sigma_R^{EXP}$ (mb) Refs.[2,23] | 1750 | | 1405 | 1270 | 1180 | 1230 | |

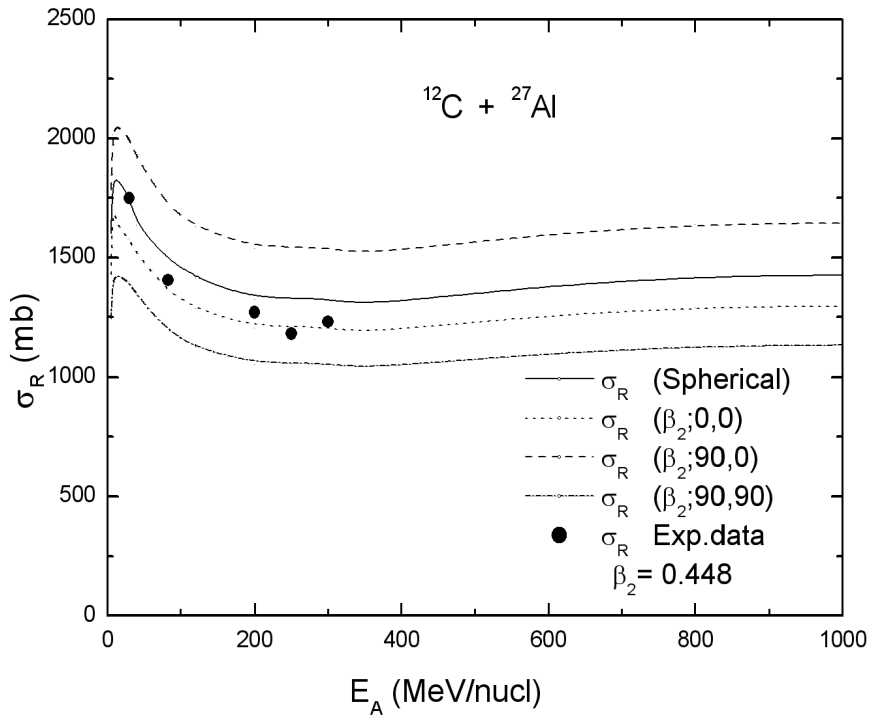

**FIG.2.** The Coulomb-modified $^{12}C + ^{27}Al$ reaction cross-sections with in-medium effect compared with experimental data [2,23] with different orientations.

From figure 2, we notice that the experimental data lies between the reaction cross-sections that are calculated with orientation $\Omega \equiv (0,0)$ and spherical ones. The results with $\sigma_R(\beta;0,0)$ are the nearer to the experimental data as shown in figure.

**Table [4]:** The reaction cross-sections for $^{20}Ne + ^{27}Al$ at different energies, using two formulas Refs. [2, 3] compared with experimental data in Refs. [2,9].

| $E_A$ (MeV/nucl) | 30 | 40 | 83 | 100 | 200 | 300 | 900 |
|---|---|---|---|---|---|---|---|
| $\sigma_R$ (mb) Ref.[2] | 2030 | 1960 | 1722 | 1662 | 1554 | 1540 | 1673 |
| $\sigma_R$ (mb) Ref.[3] | 2089 | 1837 | 1661 | 1476 | 1492 | 1498 | 1661 |
| $\sigma_R^{EXP}$ (mb) Refs.[2,9] | 2130 | | | 1465 | | 1346 | |

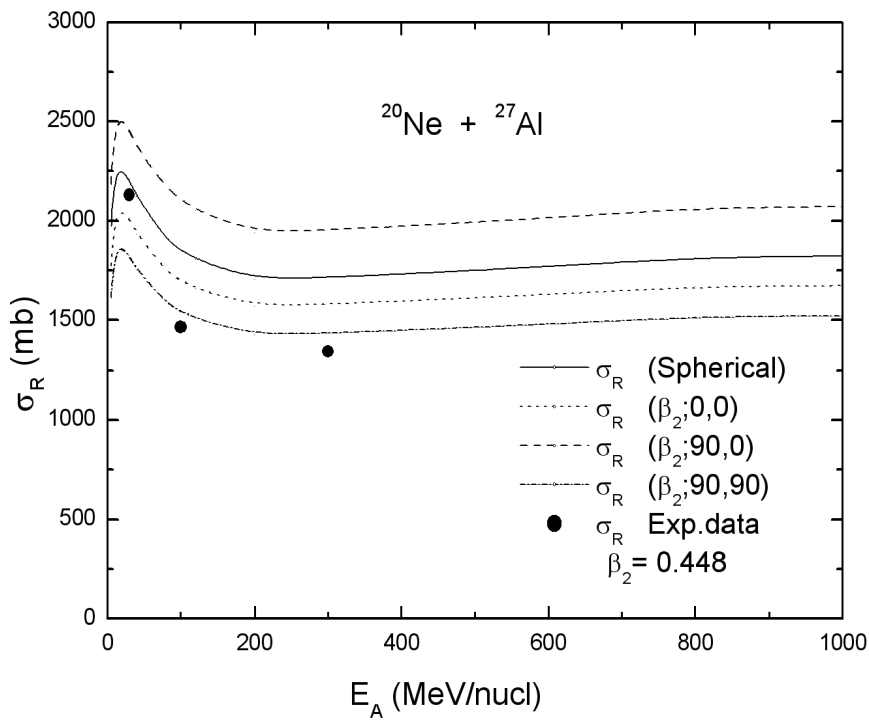

**FIG.3.** The same as Fig.1, but for $^{20}Ne + ^{27}Al$ compared with experimental data. Refs. [2,9].

The same calculations are done for the $^{20}Ne+^{27}Al$ interaction at energy range from 10 to 1000 MeV/nucl, using the prolate quadrupole deformation parameter $\beta_2$=0.448 table [2] (see Fig. 3). The experimental data are taken from [2, 9] and listed in table [4]. We notice that the reactions cross-sections that are calculated using Ref. [2] differ slightly from the ones that are calculated in Ref. [3], this result has been noticed before in reaction $^{12}C+^{27}Al$. In this case, we notice that the results with $\sigma_R(\beta;90,90)$ are the nearer to the experimental data.

**Table [5]:** The reaction cross-sections for $^{12}C+^{64}Zn$ at different energies, using two formulae Refs. [2,3] compared with experimental data in Refs. [2,9].

| $E_A$ (MeV/nucl) | 30 | 40 | 83 | 200 | 250 | 300 | 900 |
|---|---|---|---|---|---|---|---|
| $\sigma_R$ (mb) Ref.[2] | 2280 | 2185 | 1920 | 1740 | 1730 | 1720 | 1868 |
| $\sigma_R$ (mb) Ref.[3] | 2363 | 2122 | 1968 | 1795 | 1801 | 1809 | 1905 |
| $\sigma_R^{EXP}$ (mb) Refs.[2,9] | 1900 ±140 |  | 1935 | 1750 | 1735 | 1722 |  |

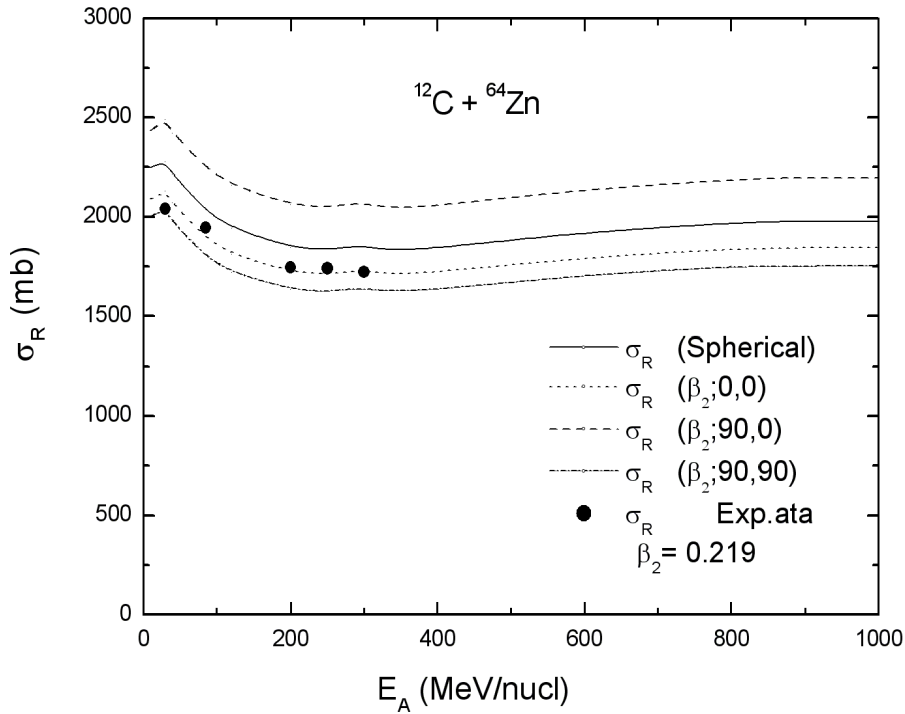

**FIG.4.** The same as Fig.1, but for $^{12}C+^{64}Zn$ compared with experimental data. Refs. [2,9].

Similarly the same calculations are done for the $^{12}C+^{64}Zn$ interaction at energy range from 10 to 1000 MeV/nucl, using the prolate quadrupole deformation parameter $\beta_2$=0.219 (see Fig.4 and table [5]). Nearly, we get the same results as reaction $^{12}C+^{27}Al$ with deformation and in-medium effect. The experimental data are near to the calculation that has been performed with deformation and orientation $\Omega \equiv (0,0)$. The reaction cross-sections are found to be largest for $\Omega \equiv (\frac{\pi}{2},0)$, while smallest for $\Omega \equiv (\frac{\pi}{2},\frac{\pi}{2})$ and the spherical reaction cross-section lies between them as given in [29].

**Table [6]:** The reaction cross-sections for $^{12}C+^{90}Zr$ at different intermediate energies, experimental data in Refs. [5, 22].

| $E_A$ (MeV/nucl) | 10 | 15 | 25 | 28.7 | 35 |
|---|---|---|---|---|---|
| $\sigma_R^{EXP}$ (mb) Refs.[5,22] | 2219 | 2297 | 2415 | 2555 | 2528 |

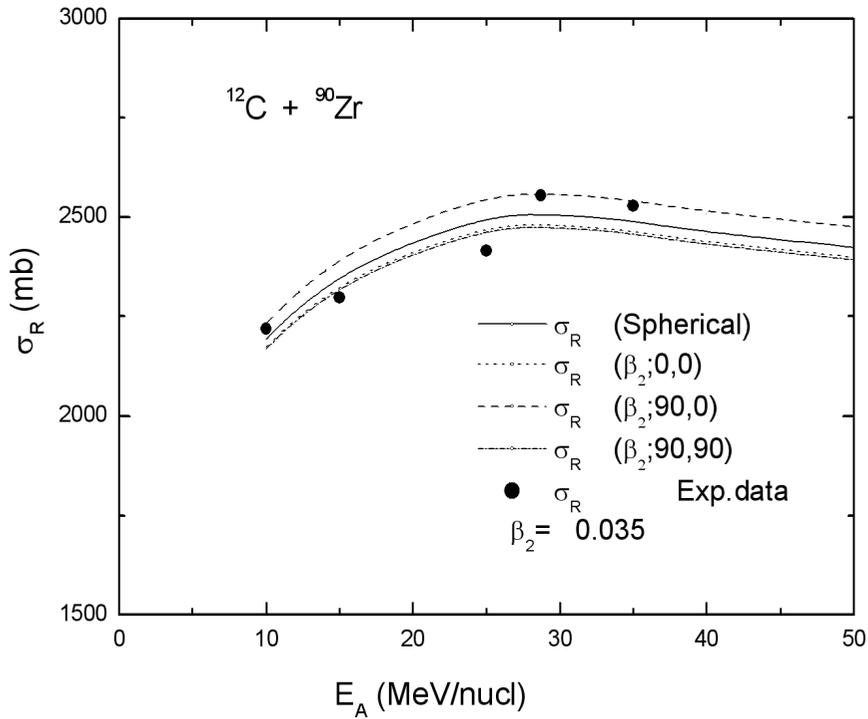

**FIG.5.** The same as Fig.1, but for $^{12}C+^{90}Zr$, compared with experimental data. Refs. [5,22].

The result of the reaction $^{12}C + {}^{90}Zr$ in the energy range from 0 to 50 MeV/nucl., are shown in Fig. 5 and table [6]. We notice that the different cases considered are near to experimental data.

**Table [7]:** The reaction cross-sections for $^{20}Ne + {}^{235}U$ at different energies, using two empirical formulae Refs. [2, 3].

| $E_A$ (MeV/nucl) | 30 | 40 | 83 | 200 | 300 | 900 |
|---|---|---|---|---|---|---|
| $\sigma_R$ (mb) Ref.[2] | 4076 | 4109 | 3940 | 3809 | 3821 | 4109 |
| $\sigma_R$ (mb) Ref.[3] | 4140 | 4007 | 4051 | 3925 | 3966 | 4140 |

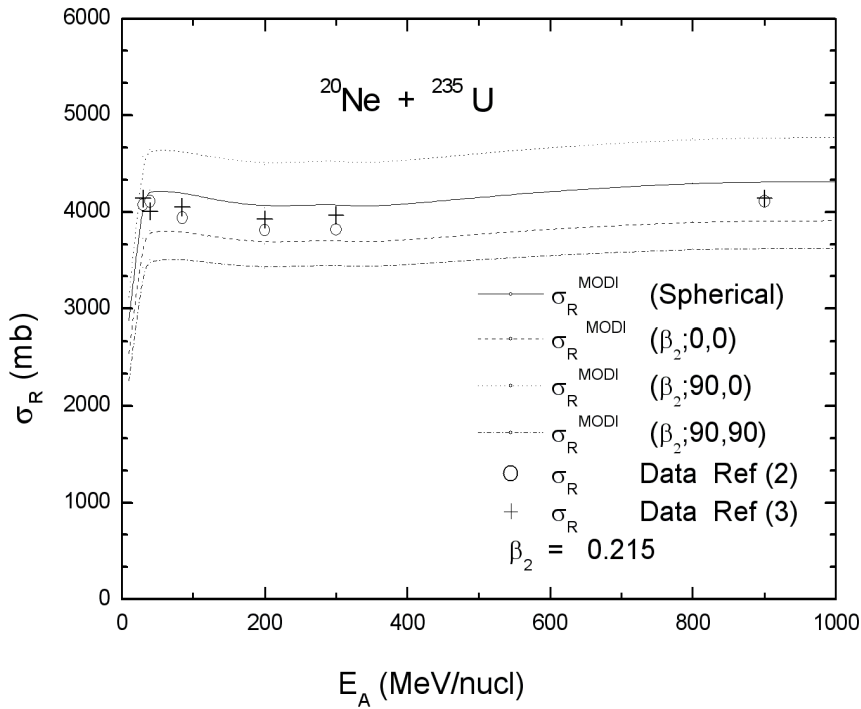

**FIG.6.** The same as Fig. 1 But for $^{20}Ne + {}^{235}U$ and compared with empirical formulas [2,3]

The reaction cross-sections for $^{20}Ne + {}^{235}U$ interaction is shown in Fig. 6. The results of both formulas are listed in table [7]. Most predictions lie between the spherical and deformed reaction cross-section with orientation $\Omega = (0,0)$.

**Table [8]:** The reaction cross-sections for $^{40}Ar + ^{238}U$ at different energies, using two empirical formulas Refs. [2,3].

| $E_A$ (MeV/nucl) | 30 | 40 | 83 | 200 | 300 | 900 |
|---|---|---|---|---|---|---|
| $\sigma_R$ (mb) Ref.[2] | 5028 | 5040 | 4815 | 4605 | 4640 | 4730 |
| $\sigma_R$ (mb) Ref.[3] | 5267 | 5107 | 5153 | 5010 | 5058 | 5255 |

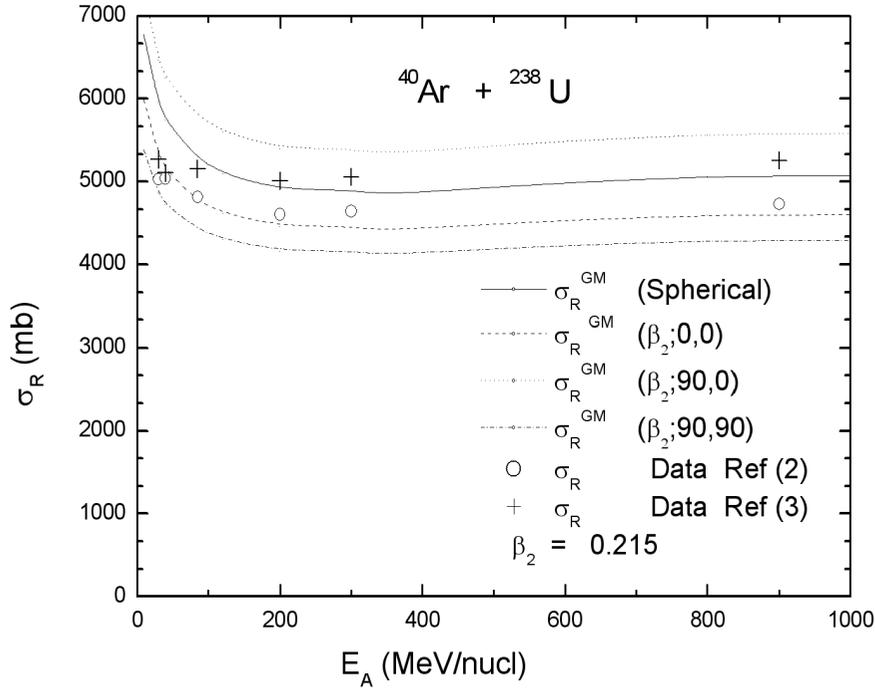

**FIG.7.** The same as Fig. 1 But for $^{40}Ar + ^{238}U$ and compared with empirical formulas [2,3]

The same calculations are done for $^{40}Ar + ^{238}U$ (see Fig.7 and table [8]) with quadrupole deformation parameter $\beta_2$=0.215 and in-medium effect, at energy range from 10 to 1000 MeV/nucl In this reaction, we notice that the reaction cross-sections calculated with orientation $\Omega \equiv (0,0)$ are in reasonable agreement with empirical formula [2]. While, the results that calculated with empirical formula [3] are nearly equal to that corresponding to the spherical reaction cross-sections

# [4] Discussions and conclusions:

The reaction cross-sections for scattering of $^{12}C$, $^{20}Ne$ and $^{40}Ar$ as projectiles on $^{235}U$, $^{238}U$, $^{27}Al$, $^{64}Zn$ and $^{90}Zr$ with energy between 10-1000 MeV/nucleon have been calculated. In general the reaction cross-sections that are calculated with the modified Glauber model I with deformation parameter $\beta_2$ and in-medium effect, give reasonable agreement with experimental data. The reaction cross-sections are found to be largest for $\Omega \equiv (\frac{\pi}{2}, 0)$, while smallest for $\Omega \equiv (\frac{\pi}{2}, \frac{\pi}{2})$. In different reactions at different energies, the experimental data agree with one of the orientation angles of the deformed nuclei. As an example, all of $^{12}C + ^{27}Al$ and $^{12}C + ^{64}Zn$ interactions agreed with $\Omega \equiv (0,0)$ and $^{20}Ne + ^{27}Al$ agreed with $\Omega \equiv (\frac{\pi}{2}, \frac{\pi}{2})$ at deformed cases. In general, both of the Coulomb and in-medium effects beside the deformation give reasonable fitting with the experimental data as shown in $^{12}C + ^{90}Zr$ interaction. For example, at high energies, the in-medium effect becomes meaningless for energy above 50 MeV/nucl according to Eq. (21). But, the modified Glauber I still presents good fitting with the experimental data for energy less than 300 MeV/nucl. The deformation effect gives better results for large energy.

Moreover, the deformed reaction cross-sections that are calculated in case of $^{20}Ne + ^{235}U$ and $^{40}Ar + ^{238}U$ interaction are in reasonable agreement with the empirical reaction cross-sections that are calculated with the Eq. (3) in Ref. [2] and Eq. (13) in Ref. [3]. Finally, the deformation effect enhances the agreement with the experimental data or other empirical parameterization formulae, at high energies.

# Reference:-